# Magnetic properties of Gd$^{3+}$ ions in the spatially distributed DNA molecules

## V.N. Nikiforov[a], V.D. Kuznetsov[b], Yu.M. Yevdokimov[c], V.Yu. Irkhin[d],*

[a]*Moscow State University, Physics Department, Low Temperatures Lab, Moscow, 119992, Russia*
[b]*Mendeleev University of Chemical Technology, Miusskaja pl. 9, Moscow, 125047, Russia*
[c]*Engelhardt Institute of Molecular Biology RAS, Vavilova str. 32, Moscow, 119991, Russia*
[d]*Institute of Metal Physics, Ural Division of RAS, Ekaterinburg, 620990, Russia*

## Abstract

Magnetic properties of DNA liquid-crystal dispersions are experimentally investigated by SQUID magnetometer. The magnetic susceptibility of pure DNA and DNA doped by La, and Gd is measured in the region 4.2–300 K. The total magnetic moment is represented as a sum of: the paramagnetic part and negative diamagnetic part. The number of paramagnetic Gd$^{3+}$ ions is calculated in a good agreement with the number of phosphate complexes. The temperature dependence of magnetic susceptibility indicates the presence of interaction between Gd$^{3+}$ magnetic moments, which is discussed in terms of long-range RKKY-type exchange in one-dimensional metals.

## 1. Introduction

DNA which is the base of genome coding provides an interesting example of one-dimensional metal, although the issue remains somewhat contradictory [1,2]. Last time, transport and magnetic properties of DNA doped by 3d-ions have been extensively investigated [3-5]. In a number of works [6-9], doping of DNA by Gd$^{3+}$ ions was reported.

$^{157}$Gd is a potential perspective agent for neutron capture cancer therapy (NCT). The microdistribution of Gd in cultured human glioblastoma cells exposed to Gd-diethylenetriaminepentaacetic acid (Gd-DTPA) was observed [10]. The Gd-DTPA penetrates the plasma membrane, and no deleterious effect on cell survival were observed. An analysis revealed a higher Gd accumulation in cell nuclei as compared with cytoplasm. This is significant for prospective NCT because the proximity of Gd to DNA increases the cell-killing potential of the short-range high-energy electrons emitted during the neutron capture reaction. Gd-containing cells bombarded by the thermal neutrons demonstrates reaction in inducing cell death. Gadolinium neutron capture therapy (Gd-NCT) utilizes the following nuclear capture reaction (NCR) of non-radioactive $^{157}$Gd by thermal neutron irradiation: $^{157}$Gd + n$_{th}$ → $^{158}$Gd + γ-rays + internal conversion electrons → Auger electrons + characteristic X-rays.

\* *E-mail address:* Valentin.Irkhin@imp.uran.ru

The success of Gd-NCT depends on a high accumulation of Gd in the tumor. Therefore, the first problem here is that a sufficient concentration of gadolinium should be retained in the tumor tissue during neutron irradiation after intratumoral injection. The second problem is the toxicity of free gadolinium to tissues [11,12]. To avoid the toxicity, it is necessary to link the $^{157}$Gd-ion to a biopolymeric molecules. Therefore, Gd complexes that can be efficiently accumulated in tumor have been sought. The chelate complexes of gadolinium (nanoparticles of various origin) have attracted considerable attention as potential molecular constructions for targeting a site and controlled gadolinium release in the tumor tissue during a Gd-NCT trial.

In the present investigation, highly rare-earth loaded double-stranded DNA liquid-crystalline particles were prepared as a potential platform for Gd-NCT. We describe in detail the results on magnetic properties of the Gd-doped DNA, which seem to be instructive for understanding the microscopic structure of this system. We propose also a new magnetic method for evaluation of the Gd content in DNA–Gd construction.

## 2. Experimental, materials and methods

We used low molecular weight double-stranded DNA molecules obtained via ultrasound cutting of DNA molecules derived from calf thymus. Samples are agglomerates of the liquid crystal dispersion DNA stabilized in an aqueous saline solution with polyethylene glycol. Thus a sample is a solution which contains the particles with the size of the order of 1000 Å, which contain ordered DNA molecules. Ordering of DNA is achieved by the change of the ionic strength of the solution, i.e., the NaCl concentration. In general, the condensation of the DNA molecules can occur with the formation of different phases: isotropic cholesteric, hexagonal and orthorhombic. In the samples presented in this work, the DNA molecules form a cholesteric phase. This phase has a number of interesting properties, including the formation of a rigid spatial structure. Therefore it is possible to determine the presence and characteristics of the structure by means of optical methods. Particles with cholesteric dispersion exist only within a certain range of osmotic pressure of the solvent defined by, e.g., polyethylene glycol concentration in the solution. Going beyond the lower limit leads to an isotropic state, going beyond the upper limit to a hexagonal packing of DNA molecules in the particle dispersion.

The molecular weight of the DNA after ultrasonic depolymerization as determined by electrophoresis in a 1% agarose gel was (0.5-0.8) 106 Da (1 Da = 1 atomic mass unit = 1.66053892 $10^{-24}$ g). XRD data obtained for the precursor particles cholesteric liquid crystal dispersion (CLCD) DNA show that the DNA concentration of the particles is close to 400 mg/ml. Further details are presented in Ref. [9].

Constructions of DNA with La and DNA with Gd were synthesized from pure DNA by $LaCl_3$ and $GdCl_3$ water solution treatment correspondingly. The 99.99% purity $GdCl_3$ was used as a source of gadolinium. At magnetic measurements, the signal from the water solutions was by 3 orders of magnitude smaller than that from DNA samples. This indicates that all the Gd ions are bound to the DNA molecules (the number of Gd ions in aqueous $GdCl_3$ solution is negligible).

The presence of the cholesteric phase in the sample was detected by means of a circular dichroism spectrum. X-ray analysis confirms presence of the ordered phase in the sample and gives DNA concentration value corresponding to the cholesteric phase. The cholesteric structure was preserved in the presence of $Gd^{3+}$, fixed in the spatial structure of the particles. Mass of each sample was 4 mg.

The temperature dependence of magnetic moment in liquid-crystal dispersions was experimentally measured by SQUID magnetometer [13]. The total magnetic moment is represented by two parts, i.e. the positive part, caused by magnetic $Gd^{3+}$ ions, magnetic residues (terminal residues etc.), and the negative diamagnetic part caused by the presence of the "rest" water (and $La^{3+}$ ions in DNA in the La-doped samples). An accuracy of SQUID magnetometer (about $10^{-10}$ emu) supports the conclusion that paramagnetism of "pure" DNA is relatively close to the case of DNA with La.

### 3. Results

The contributions to the total magnetic moment have been separated by processing of experimental temperature dependence of magnetic susceptibility. The Curie-Weiss equation was used with constant value of $\chi_0$ , namely $\chi = \chi_0 + C/(T - \Theta)$ , where $C$ is Curie constant, $\Theta$ paramagnetic Curie temperature. The Curie constant $C$ equals $N^* \cdot \mu^2/3k$, where $\mu$ is effective magnetic moment per magnetic ion, $k$ is Boltzmann constant. The number of paramagnetic centers $N^*$, i.e. of $Gd^{3+}$ ions, was calculated, the value of $\mu$ for paramagnetic center being obtained from the expression $\mu = g\mu_B \cdot [J(J+1)]^{1/2}$ as well. Results of fitting magnetic susceptibility (the calculated value of $\chi_0$, $C$ and $\Theta$) of pure DNA and DNA doped with Gd and La are presented in Table 1. The sample 5 corresponds to maximal possible $N^*$ according to the number of phosphate groups. For the chemically identical La-doped sample, it is reasonable to assume the same $N^*$ value.

Table 1. The number of La or $Gd^{3+}$ ions per DNA turn, $N^*$, and parameters of magnetic susceptibility for pure DNA, and DNA doped with Cu (according to earlier measurements [14]), La and Gd (samples 1-5).

|  | $N^*$ | $\Theta$, K | $C$, K emu/Oe g | $\chi_0$, emu/Oe g |
|---|---|---|---|---|
| DNA | 0 | −0.3 | $7.4 \ 10^{-5}$ | $+3.6 \cdot 10^{-6}$ |
| DNA Cu |  | −0.75 | $3.5 \ 10^{-4}$ | $−9.4 \ 10^{-6}$ |
| DNA La | 44 | −2 | $1.34 \ 10^{-4}$ | $+1.5 \ 10^{-6}$ |
| 1DNA Gd | 1.8 | +3.1 | $2.10 \ 10^{-3}$ | $−0.8 \ 10^{-5}$ |

| 2DNA Gd | 8.9 | +8.0 | 0.010 | −2.0 10⁻⁵ |
|---|---|---|---|---|
| 3DNA Gd | 21 | +9.0 | 0.024 | −1.7 10⁻⁵ |
| 4DNA Gd | 30 | +12.5 | 0.035 | −0.8 10⁻⁶ |
| 5DNA Gd | 44 | +3.4 | 0.053 | +1.4 10⁻⁵ |

The number of paramagnetic centers N* was calculated from the temperature dependence $\chi(T)$ in the range 77–300 K. We used ZFC and FC parts of data in this range. The value of the effective magnetic moment was estimated from ESR experiments (*g*-factor at ambient temperature makes up 1.99). Then it is possible to evaluate the number of $Gd^{3+}$ ions per gram of the sample: in the $4f^7$-state $N(Gd(4f^7)) = 3kC/(\mu_{eff})^2$, where $\mu_{eff} = g\mu_B[J(J+1)]^{1/2} = 7.9$ $\mu_B$ is the effective magnetic moment of one $Gd^{3+}$ ion

In particular, for the sample 3 ($C = 0.024$ emu/Oe g, $N(Gd(4f^7)) = 1.86 \cdot 10^{21}$ g⁻¹), the value of mass of one DNA molecule of as $m_{DNA}$ is about $8 \cdot 10^5$ Daltons, or $m_{DNA} = 1.34 \cdot 10^{-18}$ g. The concentration of DNA in the solution is $c_{DNA} = 0.00005$ g/см³; the volume of the solution, from which a sample of DNA nanoconstruction was formed, is 80 см³. According to these data, the total mass of DNA, $M_{DNA}$, is 0.004 g. This means that the number of DNA molecules in the sample is $N_{DNA} = M_{DNA}/m_{DNA} = 2.98 \cdot 10^{15}$. The number of $Gd^{3+}$ ions in the whole sample is $N^*(Gd(4f^7)) = N(Gd(4f^7))$ $M_{DNA} = 7.45 \cdot 10^{18}$. Hence, each DNA molecule contains approximately $N^*(Gd(4f^7))/N_{DNA} = 2490$ $Gd^{3+}$ ions in $4f^7$-state. Since the number of helical turns in the DNA molecule is equal to $8 \cdot 10^5/6.6 \cdot 10^3 = 120$, at each turn of DNA helix $2490/120 = 21$ $Gd^{3+}$ ions are located.

Magnetic moment of DNA with Gd includes several contributions to the total magnetic moment. They were separated by mathematical processing of experimental temperature dependence $\chi(T)$ using the Curie-Weiss equation with constant value of $\chi_0$. The $\chi_0$ value was also calculated from the high-temperature behavior of magnetic susceptibility curve with close results. Using this value of $\chi_0$ one can evaluate temperature dependence of $1/(\chi - \chi_0)$. Linear approximation of this dependence gives paramagnetic Curie temperature values (Figs. 1-3). The variation of $\chi_0$ (details of the fitting procedure) modifies somewhat the results, but not dramatically, and does not change qualitative conclusions (in particular, appreciable positive $\Theta$ for Gd-doped DNA).

For pure DNA, the signal from the sample has paramagnetic character ($\chi_0 > 0$), the Curie-Weiss contribution being small. The paramagnetism can be explained by the presence of the terminal OH groups with unsaturated oxygen bonds. Negative $\chi_0$ value in the Cu-doped sample is owing to strong diamagnetism of $Cu^{2+}$ ions. The smaller (in particular with "pure" DNA) paramagnetic contribution to $\chi_0$ in La-doped DNA is likely owing to diamagnetic $La^{3+}$ ions. Besides that, La ions can interact with magnetic residuals (phosphate complexes, terminal residuals and so on) and "heal" the magnetic radicals. The same situation can occur for Gd ions where the dependence of $\chi_0$ on Gd concentration is non-monotonous. For large Gd concentration, estimated value of $\chi_0$ is positive, which can be explained by prevalence of Van-Vleck contribution to magnetic susceptibility in comparison with the diamagnetic contribution.

The dependences $\chi(T)$ for pure and La-doped DNA are shown in Fig.1

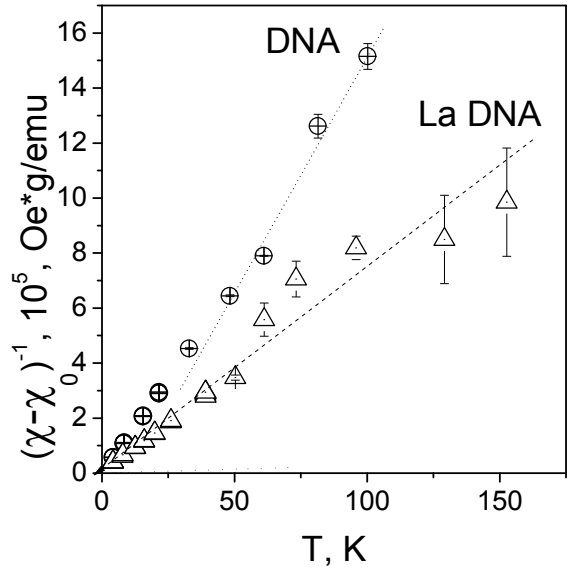

Fig. 1 Temperature dependence of magnetic susceptibility of pure DNA (circles), and La-doped DNA (triangles) at the field 526 Oe.

Gd-doped DNA samples possess much larger magnetic moments and $\chi(T)$ values as compared to pure or La-doped DNA and to measured earlier [9] Cu-doped DNA. The overall dependence $\chi(T)$ is shown in Fig.2, and details of low-temperature behavior in Fig.3. One can see occurrence of curvature in the dependence $1/(\chi(T) - \chi_0)$ while the temperature changes, which demonstrates an interaction between magnetic ions.

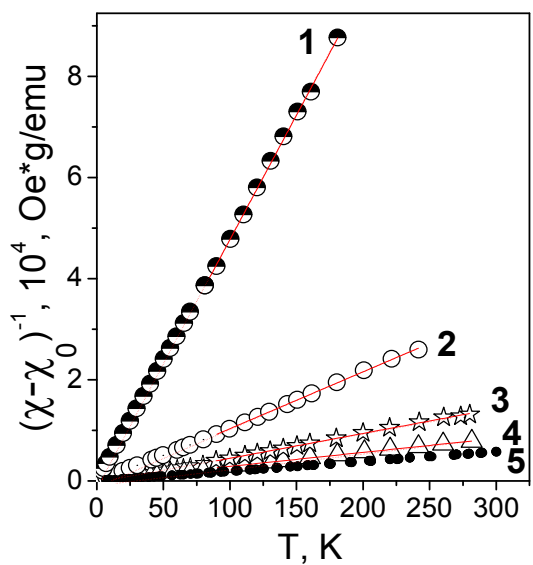

Fig. 2 Temperature dependence of magnetic susceptibility for five samples of Gd-doped DNA in a wide temperature region

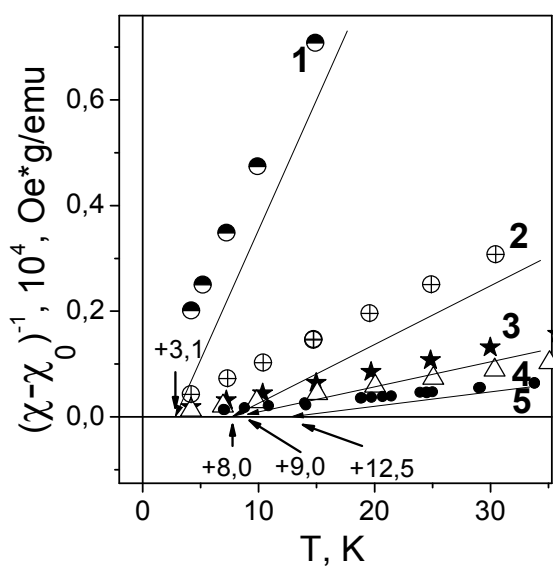

Fig. 3 Temperature dependence of magnetic susceptibility for the samples of Gd-doped DNA at low temperatures.

The prehistory of cooling down to liquid nitrogen temperatures does not affect results of magnetic experiment. Nevertheless, cooling to helium temperatures gives difference in sign of $\chi_0$, if we cool in zero magnetic field (ZFC) or nonzero (FC) field [15]. This tendency to "spin-glass" or frustration rearrangement in Gd DNA samples at low temperatures was not detected in Cu-doped DNA samples [14] with rigid nanobridges between DNA molecules. This fact illustrates that the molecules of DNA with Gd have some degrees of freedom. Liquid crystal dispersion (LCD) of DNA with Gd is relatively not rigid. Thus it is reasonable to assume that $Gd^{3+}$ ions located on the phosphate $P^{2+}$ group are influenced by intra-helix and extra-helix Coulomb and magnetic interactions.

## 4. Discussion and conclusions

According to geometry of a DNA molecule (the step distance between protein bases is 3.4 Å), the distance between atoms Gd on phosphates groups will be about 5-10 Å. At the same time, appreciable values of paramagnetic Curie temperature $\Theta$ indicate considerable interaction between Gd magnetic moments (no crystal field influence is expected for $Gd^{3+}$ ions). With further increase of Gd concentration (sample 5) the value of $\Theta$ can fall, which may mean destruction of conductivity in DNA chains. Thus the occurrence of appreciable $\Theta$ is unlikely to be due to merging of close Gd atoms.

Usual exchange interactions (like superexchange in insulators) between Gd ions can hardly work at the large separations. On the other hand, the RKKY interaction (characteristic for metals) can operate, being especially long-range in the one-dimensional case. Indeed, in this case the exchange parameter is expressed in terms of the sine integral function $Si(2k_F R)$ [16,17] and behaves at large distances $R$ as $J^{RKKY}(R) \sim \cos(2k_F R)/R$ (instead of $1/R^3$ in the three-dimensional case and $1/R$ in the two-dimensional case). Thus the RKKY interaction can occur for conducting DNA molecules. The interchain magnetic exchange interaction seems to be less significant, because the whole turn of the helix is equal 34 Å, so that nearest Gd ion on other helix (second chain) will be chaotically located on rather long distances.

Recently, the mechanism of DNA conductivity was discussed using a theoretical approach based on the results of electron localization in correlated disordered potentials [18,19]. Since most of the mutations in DNA are successfully healed, one may assume the existence of charge transport through delocalized states that are responsible for the transfer of information at long distances. According to Ref. [18]. the exons (the parts where the genetic information is written) have narrow bands of extended (practically delocalized) states, unlike the introns (the parts without apparent information for protein synthesis) where all the states are well localized. Thus the localization length and the conductance of a given segment of the DNA molecule are directly related to the genetic information stored in this segment.

By analogy with the information transfer, we can assume that a long-range exchange between magnetic moments via current carriers may occur, although we cannot indicate precisely the mechanism of mediation of interaction (in particular, RKKY exchange, as for usual metals, or double exchange in narrow bands).

To conclude, DNA chains with Gd ions located on the helix spatially ordered phosphate complexes yield a bright example of one-dimensional metal with interacting magnetic moments. This assumption is supported by finite value of paramagnetic Curie temperature and the difference between ZFC and FC magnetic moment measurements in DNA with Gd at low temperatures

## Acknowledgements

We are grateful to B. L. Oksengendler and Yu. N. Shvachko for useful discussions. This work is supported in part by the Program of fundamental research of RAS Presidium "Quantum mesoscopic and disordered structures", project No. 12-P-2-1041.